\documentclass[reprint,amsmath,amssymb,
aps,prl,twocolumn,superscriptaddress,groupedaddress]{revtex4}

\usepackage{graphicx}
\usepackage{dcolumn}
\usepackage{bm}
\usepackage{float}
\usepackage{amssymb}

\begin{document}

\title{Exotic Superconducting Properties in Single Crystals of Topological Nodal Semimetal PbTaSe$_2$}

\author{Cheng-Long Zhang}
\affiliation{International Center for Quantum Materials, School of Physics, Peking University, Beijing 100871, China}

\author{Zhujun Yuan}
\affiliation{International Center for Quantum Materials, School of Physics, Peking University, Beijing 100871, China}

\author{Guang Bian}
\affiliation{Joseph Henry Laboratory, Department of Physics, Princeton University, Princeton, New Jersey 08544, USA}

\author{Su-Yang Xu}
\affiliation{Joseph Henry Laboratory, Department of Physics, Princeton University, Princeton, New Jersey 08544, USA}

\author{Xiao Zhang}
\affiliation{International Center for Quantum Materials, School of Physics, Peking University, Beijing 100871, China}

\author{M. Zahid Hasan}
\affiliation{Joseph Henry Laboratory, Department of Physics, Princeton University, Princeton, New Jersey 08544, USA}

\author{Shuang Jia\footnote{Corresponding author: gwljiashuang@pku.edu.cn}}
\affiliation{International Center for Quantum Materials, School of Physics, Peking University, Beijing 100871, China}
\affiliation{Collaborative Innovation Center of Quantum Matter, Beijing 100871, China}

\date{\today}

\begin{abstract}
We report electronic properties of superconductivity in single crystals of topological nodal-line semimetal PbTaSe$_2$. Resistivity, magnetic susceptibility and specific heat measurements were performed on high-quality single crystals. We observed large, temperature-dependent anisotropy in its upper critical field ($H_{c2}$).
The upper critical field measured for $H$ $\parallel$ \textbf{ab} plane shows a sudden upward feature instead of saturation at low temperatures. The specific heat measurements in magnetic fields reveal a full superconducting gap with no gapless nodes.

\end{abstract}

\pacs{Valid PACS appear here}
\maketitle

\section{\romannumeral1. INTRODUCTION}
A time-reversal invariant topological insulator possesses a symmetry-protected surface state due to its non-trivial bulk band wave function orderings \cite{TI_Hasan}.
The topological classification of the phases in terms of their band structures is crucial for the study of their topological properties. Recent discovery of Weyl semimetals \cite{TaAs_Arpes_Hasan,TaAs_Arpes_HongDing,NbAs_Arpes_Hasan,TaAs_Chenyulin,TaP_Dinghong} has shifted the focus from bulk insulators to metallic phases in which bulk band crossings occur. In contrast to Weyl semimetals whose crossing is essentially 0-dimensional (0D), nodal-line semimetals possess 1-dimensional (1D) crossing line on their band structures \cite{Nodal_semimetals_Burkov}.
Recently, angle-resolved photoemission spectroscopy (ARPES) measurements for single crystalline PbTaSe$_2$ showed the existence of bulk nodal-line band structure and non-trivial surface states \cite{PbTaSe2_ARPES1_Bianguang}. With the existence of the bulk superconductivity, proximity-induced  topological superconductivity on the non-trivial surface states which host Majorana fermions was predicted theoretically \cite{PbTaSe2_theory1_Bianguang}. The pairing symmetry of the bulk superconductivity plays a central role for realizing the topological superconductivity in the surface of PbTaSe$_2$. Bulk superconducting properties have been reported in polycrystalline PbTaSe$_2$ previously \cite{PbTaSe2_cava,PbTaSe2_upward}, but the properties of anisotropy have not been studied exclusively due to the lack of single crystals. The inner pairing mechanism for this non-central symmetric superconductor is not clear yet.

Nodal-line semimetal PbTaSe$_2$ can be viewed as Pb atoms intercalated TaSe$_2$ in which triangle lattices of Pb atoms layers are sandwiched between the hexagonal TaSe$_2$ layers. This alternately stacked structure is noncentrosymmetric (Fig. 1(a)). The parent material TaSe$_2$ has been well studied due to its charge-density wave (CDW) instability and many related anomalies in its transport properties \cite{TaSe2_CDW}. When the layers of Pb atoms are intercalated between TaSe$_2$ layers, the CDW transition is suppressed and a superconducting transition occurs at 3.8 K \cite{PbTaSe2_FangchengChou}. Here we show detailed results of magnetization, heat capacity and electrical transport measurements on high-quality single crystals of PbTaSe$_2$. We observed upward features and large anisotropy in upper critical field ($H_{c2}$) which is well below its Pauli limit. Heat capacity measurements reveal a full superconducting gap. All the properties in this noncentrosymmetric superconductor provide a clue for realizing a topological superconductor with nontrivial gapless surface states \cite{topo.SC_Xiaoliang,topo.SC_Ando}.

\section{\romannumeral2. CALCULATION AND EXPERIMENT}
PbTaSe$_2$ single crystals were prepared by chemical vapor transport (CVT) method using PbCl$_2$ as a transport agent.
Polycrystalline PbTaSe$_2$ was synthesized via a solid reaction in sealed, evacuated quartz tubes with a growth condition similar as that in Ref. \cite{PbTaSe2_cava}. All process were operated in circulated Argon box with oxygen and water content less than 0.5 ppm. About 0.45 g of PbTaSe$_2$ powder and 15 mg of PbCl$_2$ were put at one end of a long quartz tube. Then the tube was evacuated and sealed off and placed in a three-zone furnace with the source zone at 900 $^oC$ and the sink zone at 800 $^oC$ for a week. Large flake-like single crystals obtained at the sink zone have in-plane size near 3 $\times$ 3 mm but ultrathin in \textbf{c}-axis direction (Fig. 1(b)). The powder x-ray diffraction (XRD) and Rietveld refinement results for PbTaSe$_2$ confirm that the crystal structure is noncentrosymmetric with a space group of $P$$\bar{6}$$m2$ (Fig. 1(c)). The arrow in Fig. 1(c) demonstrates the characteristic XRD peak for noncentrosymmetric space group $P$$\bar{6}$$m2$, which is absent for the centrosymmetric space group $P6/mmm$.

Resistivity, heat capacity and magnetization measurements were performed in a Quantum Design physical property measurement system (PPMS-9). dc magnetization was measured in the Vibrating Sample Magnetometer (VSM) option. Heat capacity and resistivity measurements below 2 K were carried out in a dilution refrigerator in PPMS-9. A standard four-probe method for resistivity measurements was adopted with employing silver paste contacts. The contact resistance in this experiment is less than 6 $\Omega$ . ac electric currents were applied in the basal plane of the crystals. Band structure calculations were performed under the framework of the generalized gradient approximation (GGA) of density functional theory (DFT) \cite{GGA1,GGA2}, and the spin-orbit coupling (SOC) was incorporated.

\begin{figure}[!h]
\includegraphics[clip, width=0.5\textwidth]{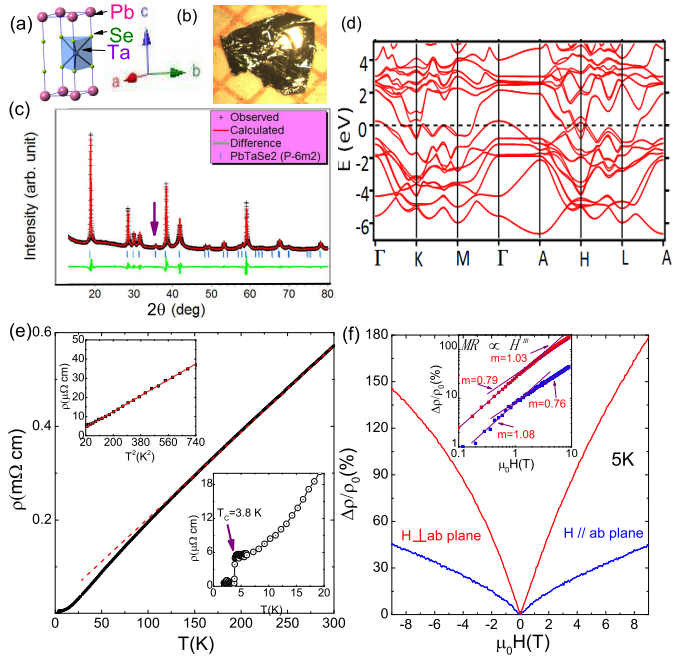}
\caption{\label{fig:epsart} (a) The crystal structure of PbTaSe$_2$. (b) A photo of as-grown single crystals of PbTaSe$_2$. The red mesh grids are in size of 1mm$\times$1mm. (c) Powder x-ray diffraction (XRD) and Rietveld refinement results for single crystalline PbTaSe$_2$. The arrow shows the characteristic XRD peak of noncentrosymmetric space group $P$$\bar{6}$$m2$. (d) First-principle band structure calculations show band inversions around $H$ point. (e) Temperature-dependent resistivity from 300 K to 2 K. The dashed red line shows linear temperature dependence of resistivity in high temperature region. Inset (upper-left): Resistivity versus $T^2$ below the Debye temperature $\Theta_D$ exhibits linear dependence. Inset (lower-right): The temperature of superconducting transition is identified as 3.8 K in zero magnetic field. (f) Magnetoresistivity in \textbf{ab} plane and out of \textbf{ab} plane at 5 K. Inset: The log-log plot shows linear MR in low fields. The data were not symmetrized.}
\end{figure}

\section{\romannumeral3. RESULTS}
Figure 1(d) presents the main features of the band structure calculations for PbTaSe$_2$. There is a giant hole pocket around $\Gamma$ point as the main contribution to the density of states (DOS) close to the Fermi level. Another contribution to DOS at the Fermi level comes from the four bands including two electron-like conduction bands and the two hole-like valence bands which cross each other near $H$ point. This band structure calculations is confirmed by ARPES measurements in Ref.\cite{PbTaSe2_ARPES1_Bianguang}.

The temperature-dependent resistivity was measured from 300 K down to 2 K in zero field (Fig. 1(e)). The samples described in this paper show large residual resistivity ratio RRR = $\rho(300 K)/\rho(5 K)=107$. But we found that the quality of crystals varies batch-to-batch due to a stacking disorder problem \cite{eppinga1980generalized}. The stacking disorder effects on this topological system will be clarified by experiments in the future. The superconducting transition temperature is identified at 3.8 K as shown in the lower-right inset of Fig. 1(e). Temperature-dependent resistivity shows a Fermi liquid behavior with a relation of $\rho=\rho_0+AT^2$ below 25 K (the upper-left inset of Fig. 1(e)). Fitting yields the residual resistivity $\rho_0$ = 3.8 $\mu\Omega$ and coefficient $A$ = 0.045 $\mu\Omega$ cm/K$^2$. High temperature resistivity ($T$\textgreater $\Theta_D$) exhibits a linear temperature dependence, indicating the dominance of electron-phonon scattering. Figure 1(f) shows results of magnetoresistance (MR = $\Delta\rho/\rho$) for $H$ $\parallel$ \textbf{ab} plane and $H$ $\perp$ \textbf{ab} plane at 5 K. The quantity analysis of field-dependent MR is carried out by replotting the data in log-log scale as shown in the inset of Fig. 1(f). The MR shows linear dependence in the fields lower than 1 T for $H$ $\parallel$ \textbf{ab} plane and 2 T for $H$ $\perp$ \textbf{ab} plane at 5 K, and then it tends to deviate from the linear relation in higher fields (Fig. 1(f)). The linear MR in low fields is probably related to the nontrivial bulk bands \cite{hu2008LinearMR}, but the exact reason for this field dependence is not clear.

\begin{figure}[!h]
\includegraphics[clip, width=0.4\textwidth]{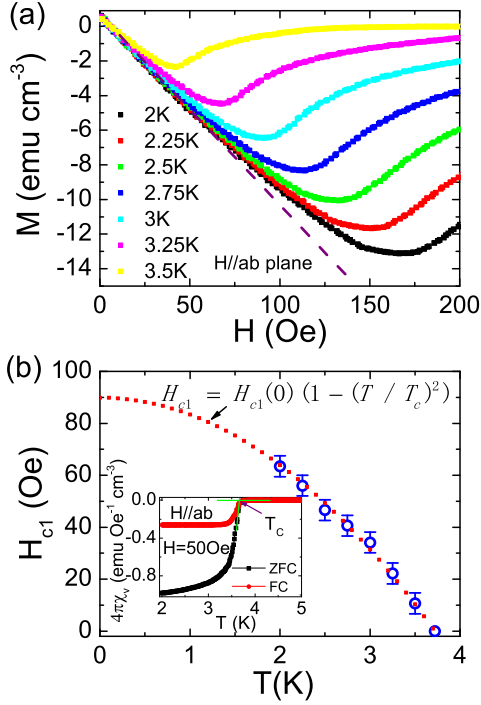}
\caption{\label{fig:epsart} (a) Low-field magnetization $M(H)$ for $H$ $\parallel$ \textbf{ab} plane at different temperatures. The dashed line is a linear fit to the data at 2 K. (b) Lower critical field versus temperature for PbTaSe$_2$. The inset shows the ZFC and FC magnetic susceptibilities measured in a constant magnetic field of 50 Oe.  }
\end{figure}

Figure 2(a) shows magnetization ($M$) versus $H$ at different temperatures below $T_c$. The values of the lower critical filed ($H_{c1}$) are determined by the points of 10\% deviation of $M$ from the linear fit to the low field data (Fig. 2(a)). Then $H_{c1}$ was plotted in Fig. 2(b) versus temperatures. Fitting to the data by using $H_{c1}(T)$ = $H_{c1}(0)[1-(T/T_c)^2]$ yields $H_{c1}(0)$ = 90 Oe. The inset of Fig. 2(b) shows that the field-cooling (FC) superconducting diamagnetic signals are much smaller than the zero-field-cooling (ZFC) signals caused by pinning of the vortices, which indicates a type-\uppercase\expandafter{\romannumeral2} superconductor, albeit its large RRR. The estimated superconducting transition temperature around 3.7 K is in good agreement with that of resistivity measurements.
\begin{figure}[!h]
\includegraphics[clip, width=0.5\textwidth]{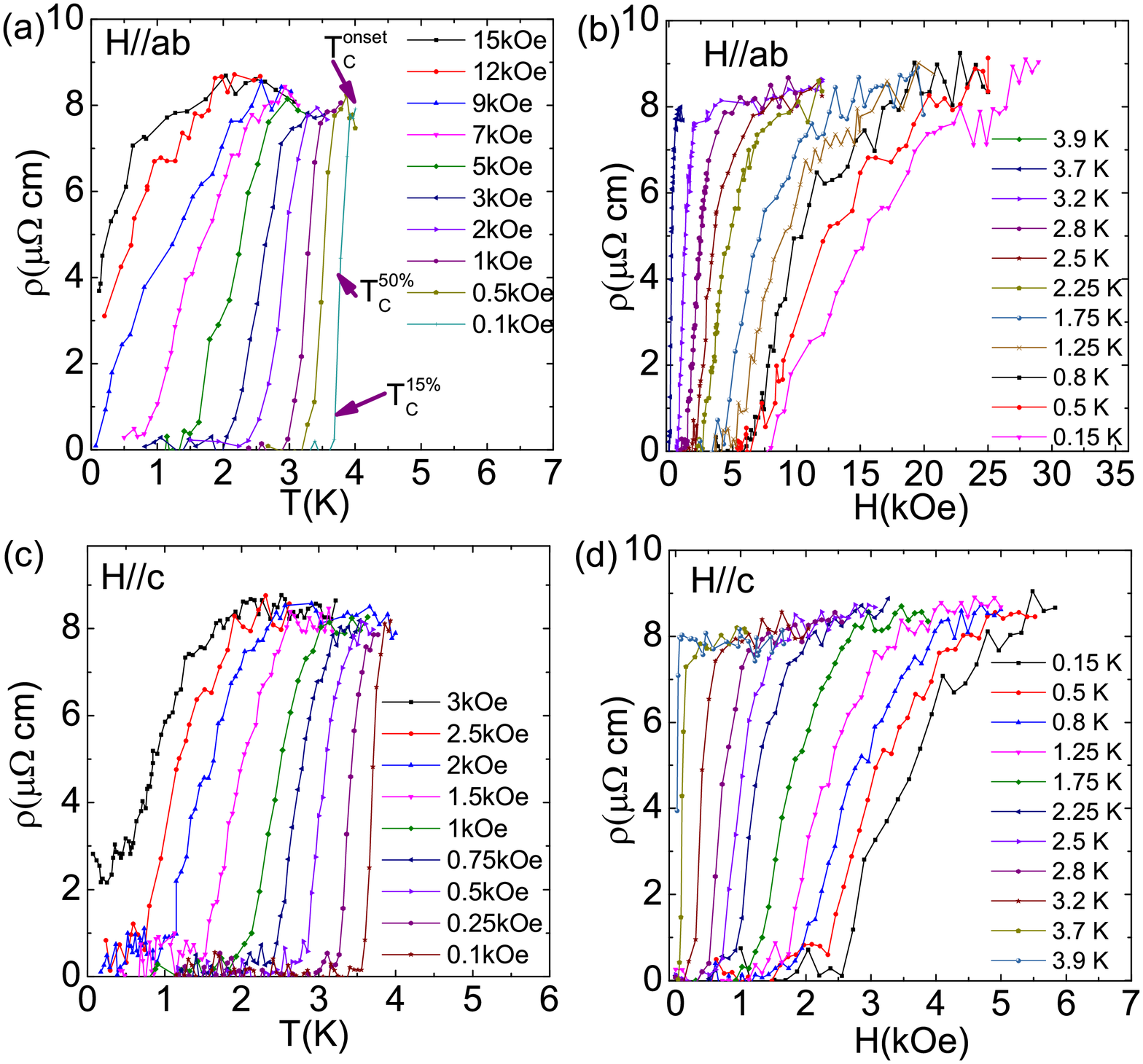}
\caption{\label{fig:epsart} Temperature-dependent resistivity in different magnetic fields and field-dependent resistivity at different temperatures for $H$ $\parallel$ \textbf{ab} plane ((a)\&(b)) and $H$ $\perp$ \textbf{ab} plane ((c)\&(d)), respectively. Three criteria illustrated in (a) are used to determine T$_c$ in each field.
  }
\end{figure}

The measured superconducting transition (Fig. 3) shows significant broadening in applied field. Three different criteria (onset, 50\% and 15\%) are subsequently adopted to determine the transition temperatures as shown in Fig. 3(a). The midpoints of the superconducting transition in resistivity are chosen as the representative data which was used in further analyze. The upper critical fields for $H$ $\parallel$ \textbf{ab} plane ($H_{c2}^{ab}$) and $H$ $\perp$ \textbf{ab} plane ($H_{c2}^{c}$) under different criterion are plotted as functions of temperature in Fig. 4(a). The upper critical fields extracted from temperature-dependent resistivity curves (open symbols) coincide with those from field-dependent resistivity curves (filled symbols). Using the equation $H_{c2}(T)$ = $H_{c2}(1-t^2)(1+t^2)$, where $t$ = $T/T_c$, according to the Ginzburg-Landau theory, we estimate $H_{c2}^{ab}(0)$ and $H_{c2}^{c}(0)$ as 8.7 kOe and 2.4 kOe, respectively. On the other hand, we obtain the initial slopes $dH_{c2}/dT\vert_{T=T_c}$ to be -0.061 T/K ($H$ $\perp$ \textbf{ab} plane) and -0.19 T/K ($H$ $\parallel$ \textbf{ab} plane). Then $H_{c2}^{ab}(0)$ and $H_{c2}^{c}(0)$ for an isotropic full superconducting gap in the clean limit and without SOC were estimated by using the equation $H_{c2}(0)$ = $-0.73(dH_{c2}/dT)\vert_{T=T_c}T_c$ from the Werthammer-Helfand-Hohenberg (WHH) model \cite{WHH0.690.73} as 5.3 kOe and 1.7 kOe, respectively. As shown in Fig. 4(a), $H_{c2}$ shows an upward feature at low temperatures. This behavior is not consistent with theoretical predictions referred above. The measured $H_{c2}(0)$ is much higher than the estimations. Figure 4(b) shows $H_{c2}$ estimated as the midpoints of heat capacity jump for $H$ $\perp$ \textbf{ab} plane (see following section), which also displays an upward feature for $T$ \textless 5 K. Because of the large resistivity fluctuations near the offset $T_c$, a 15\% criterion is adopted for determining the $T_c$. The data obtained by using the 15\% criterion from resistivity measurements are larger than the $H_{c2}$ extracted from heat capacity, this inconsistency will be remedied by the true offset criterion in further high-precision measurements. Figure 4(c) shows the temperature-dependent anisotropic ratio of upper critical fields $\gamma$ = $H_{c2}^{ab}$/$H_{c2}^c$. The highly temperature-dependent feature is similar as what had been observed in multi-band superconductor MgB$_2$ \cite{MgB2_Hc2}.

\begin{figure}[!h]
\includegraphics[clip, width=0.5\textwidth]{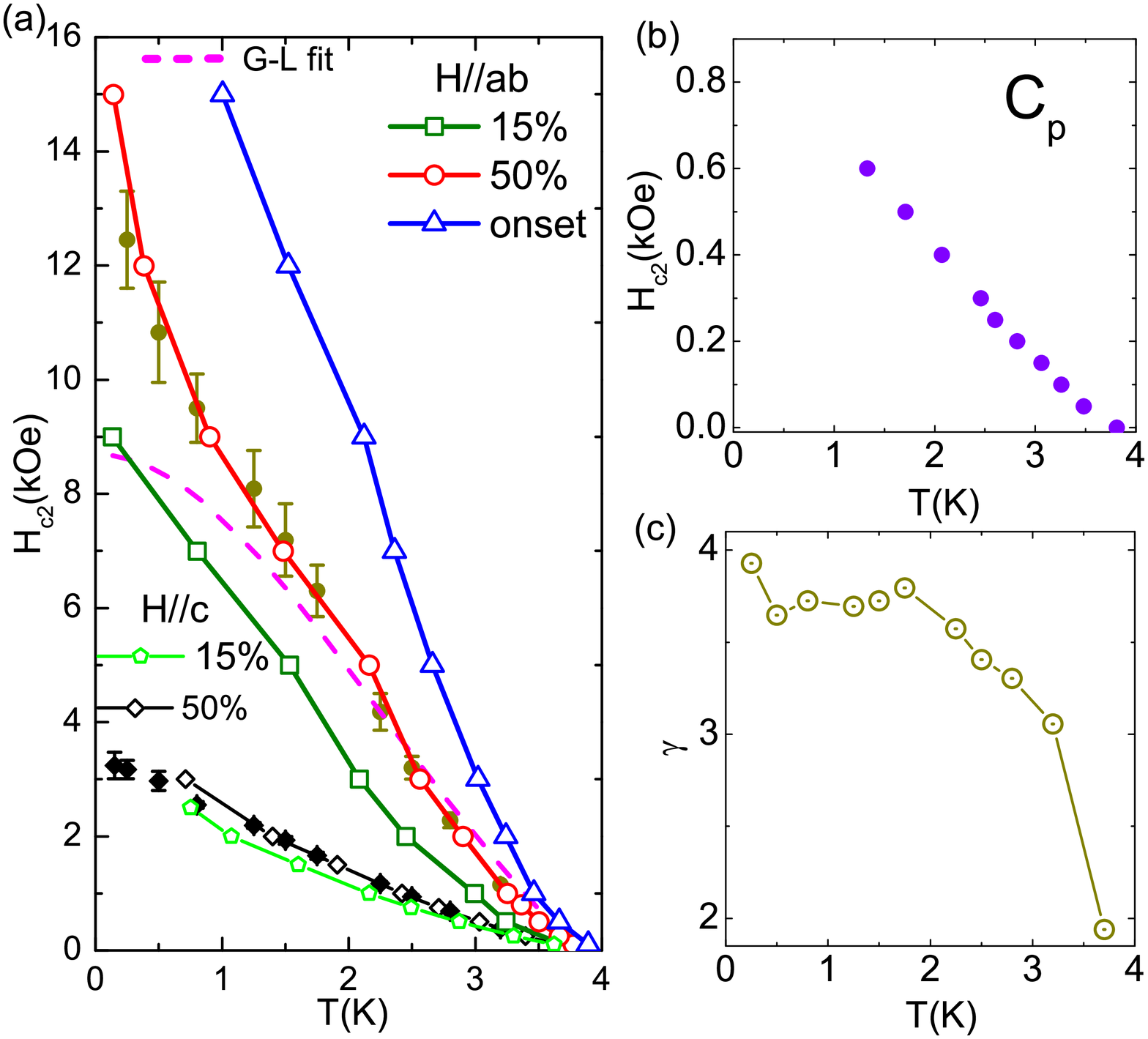}
\caption{\label{fig:epsart} (a) Extracted upper critical fields $H_{c2}(T)$ for the two field orientations versus temperature. Three criteria for determining transition temperatures were used here. Open and filled symbols represent the $H_{c2}(T)$ extracted from temperature- and field-dependent resistivity curves, respectively. The pink dashed line denotes a Ginzburg-Landau (GL) fit. (b) Upper critical fields extracted from specific heat measurements show a similar upward feature. (c) The anisotropy of upper critical fields in two orientations is denoted as $\gamma$ = $H_{c2}^{ab}$/$H_{c2}^c$.  }
\end{figure}

\begin{table}[!h]
\caption{\label{tab:table1}%
Summary of superconducting parameters of single and poly-crystalline PbTaSe$_2$
}
\begin{ruledtabular}
\begin{tabular}{lcdr}
\textrm{Parameter}&
\textrm{Unit}&
\textrm{single-} &
\textrm{poly-} \cite{PbTaSe2_cava}\\
\colrule
$T_c$ & K & 3.8 & 3.72\\
$H_{c1}(0)$ & mT & 9 & 7.5\\
$H_{c2}^{ab}(0.2)$ & T & 1.25 & 1.47\\
$H_{c2}^{c}(0.2)$ & T & 0.32 & \\
$H_{c2}^{ab}$/$H_{c2}^c(0.2)$ & T & 3.9 & \\
$\xi_{GL}(0)$ & nm & 16.2 & 15\\
$\kappa(0)$ &  &  & 17\\
$\gamma_n$ & mJ mol$^{-1}$ K$^{-2}$ & 6.01 & 6.9\\
$\Delta{C_e}/\gamma_n{T_c}$ &  & 1.71 & 1.41\\
$\mu_0H^{Pauli}$ & T & 6.9 & 6.8\\
$\Theta_D$ & K & 161 & 112\\
$\lambda_{ep}$ &  &  & 0.74\\
\end{tabular}
\end{ruledtabular}
\end{table}
\begin{figure}[!h]
\includegraphics[clip, width=0.5\textwidth]{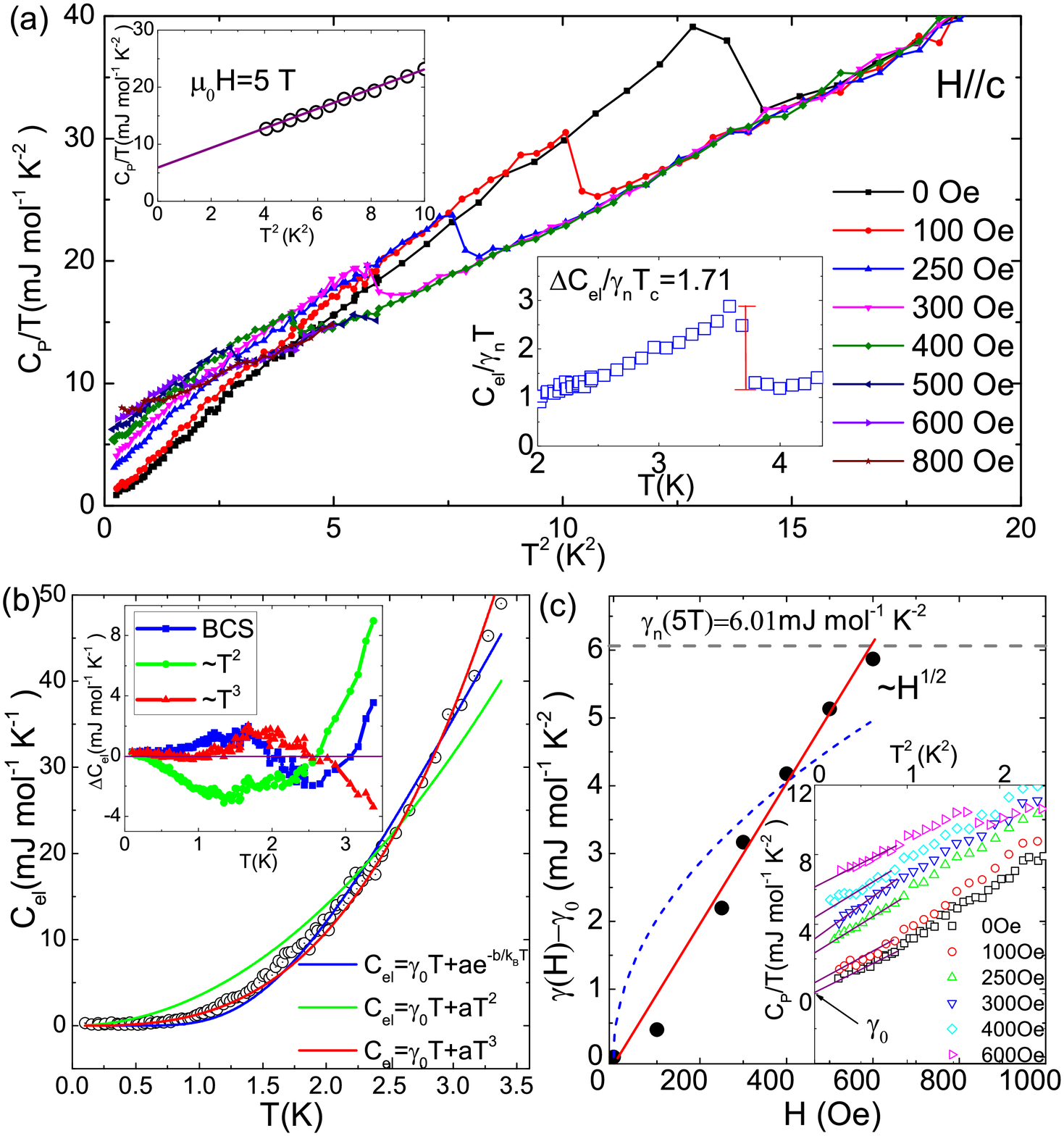}
\caption{\label{fig:wide} (a) Specific heat divided by temperature ($C_p/T$) as a function of $T^2$ in different magnetic fields along \textbf{c}-axis.
Inset (upper-left): $C_p/T$ versus $T^2$ in 5 T along \textbf{c}-axis. Inset (lower-right): Normalized electronic specific heat, $C_{el}/\gamma_nT$, as a function of temperature. (b) Temperature-dependent electronic specific heat fitted by three models. Inset: The difference between the models and raw data. (c) $\gamma(H)-\gamma_0$ is linear dependent with the magnetic field (red line). The blue dotted line shows $H^{1/2}$ dependence. Inset: $C_p/T$ as a function of $T^2$ in low temperature region. Linear extrapolations on the vertical axis give values of $\gamma(H)$. }
\end{figure}

Figure 5 shows the data and analysis of the specific heat for PbTaSe$_2$. A sharp jump in $C_p/T$ around $T_c$ = 3.6 K denotes the onset of superconductivity (Fig. 5(a)). The upper-left inset of Fig. 5(a) shows $C_p/T$ is linear with respect to $T^2$ in a small range of temperatures in $H$ = 5 T, which exceeds the $H_{c2}$ for PbTaSe$_2$. The linear fit with a formula  $C_p/T$ = $\gamma_n+\beta{T^2}$ yields the intercept $\gamma_n$ = 6.01 mJ K$^{-2}$ mol$^{-1}$ and the slope $\beta$ = 1.70 mJ K$^{-4}$ mol$^{-1}$, where $\gamma_n$ is the normal state Sommerfeld coefficient and $\beta{T^3}$ is the lattice contribution to the specific heat. Both two specific heat coefficients are slightly smaller than that of polycrystalline PbTaSe$_2$ \cite{PbTaSe2_cava}. By using the formula $\Theta_D$ = $[(12/{5\beta})\pi^4nR]^{1/3}$, where $R$ = 8.314 J mol$^{-1}$ K$^{-1}$ and $n$ = 4 for PbTaSe$_2$, the Debye temperature is estimated as $\Theta_D$ = 161 K. The electronic specific heat $C_{el}$ is estimated by subtracting the part of lattice contribution in the total specific heat. In the lower-right inset of Fig. 5(a), $C_{el}/{\gamma_nT}$ is plotted as a function of $T$. The dimensionless quantity $C_{el}/{\gamma_nT_c}$ is calculated as 1.71, larger than the value of 1.43 in weak-coupling BCS theory. This value of $C_{el}/{\gamma_nT_c}$ is also larger than that for polycrystalline PbTaSe$_2$ in Ref. \cite{PbTaSe2_cava}.

Temperature-dependent electronic specific heat ($C_{el}$) below $T_c$ is shown in Fig. 5(b). In order to see whether there are nodes on the superconducting gap, the data was fitted by three models $C_{el}$ $\propto$ $T^2$, $T^3$ and $e^{-b/{kT}}$ which are expected for line nodes, point nodes and a full gap, respectively. The value of $\gamma_0$ is near zero, thus the term $\gamma_0T$ is held constant throughout the fitting. The inset of Fig. 5(b) shows the difference between data and models. Both models of $C_{el} \propto$ $T^3$ and $\propto$ $e^{-b/{kT}}$ provide satisfying fitting and cannot be distinguished here.

The measurements of the field dependent electronic specific heat ($C_{el}(H)$) will give out the information of the low-energy excitations in superconducting states. For a highly anisotropic gap or a gap with nodes, the Sommerfeld coefficient in magnetic fields follows the relation $\gamma(H)$ $\propto$ $H^{1/2}$ \cite{volovik_H0.5}. While for a fully gapped superconductor, $\gamma(H)$ is proportional to the number of field-induced vortices, resulting in $\gamma(H)$ $\propto$ $H$. In Fig. 5(c), the $\gamma(H)$-$\gamma_0$ extracted from the inset of Fig. 5(c) exhibits a linear dependence on the magnetic field, which suggests a fully gapped superconducting state with no line nodes or point nodes. This full gap also indicates the triplet state is not dominated in the noncentrosymmetric superconductor PbTaSe$_2$.

\section{\romannumeral4. DISCUSSION AND CONCLUSION}
We finally discuss the upward feature in $H_{c2}$, which has been observed in polycrystalline samples and mainly explained by the model of two-band superconductivity \cite{PbTaSe2_cava,PbTaSe2_upward}. Our results on high-quality single crystals show that $H_{c2}$ exhibits a sudden upward feature with no saturation down to 0.15 K. An upward in $H_{c2}$ has been observed in high-$T_c$ superconductors \cite{HTSC_Upward1,HTSC_Upward2} and typical two-band superconductor MgB$_2$ \cite{MgB2_Hc2}. But we notice that the upward feature for PbTaSe$_2$ are different with these two systems. The  $H_{c2}$ for PbTaSe$_2$ at low temperatures seems to be more significant than in relatively higher temperatures, which is also different from that in MgB$_2$ \cite{MgB2_Hc2}.
The background of superconductive PbTaSe$_2$ is different from that in high-$T_c$ superconductors as following: 1. $H_{c2}$ for PbTaSe$_2$ is well below Pauli limit; 2. there is no magnetic elements in PbTaSe$_2$. As shown in our calculations and ARPES data \cite{PbTaSe2_ARPES1_Bianguang}, the band structure for PbTaSe$_2$ around Fermi level is complicated. The Fermi surface comes from multiple bands.
The origin of the upward feature of $H_{c2}$ phase diagram in this full-gap superconducting system needs further experiments to clarify.

To conclude, we report the superconducting properties for single crystalline PbTaSe$_2$ which is a topological nodal-line semimetal in its normal state. We found that it has a full superconducting gap below $T_c$. We confirmed the upward features in its $H_{c2}$, which may infer exotic underlying physics.

\begin{acknowledgments}
C.L.Z. and S.J thank T.Neupert for valuable discussions. C.L.Z., Z.J.Y. and X.Z. thank Yuan Li and Ji Feng for using their instruments. S.J. is supported by National Basic Research Program of China (Grant Nos. 2013CB921901 and 2014CB239302). The work at Princeton is supported by U.S. DOE DE-FG-02-05ER46200 and by Gordon and Betty Moore Foundation through Grant GBMF4547 (Hasan) for sample characterization.

Recently, we noticed a work (emphasized on the thermal conductivity) about single crystalline PbTaSe$_2$\cite{PbTaSe2_LiSY}.

\end{acknowledgments}

\end{document}